\documentstyle[12pt]{article}
\begin{document}
\baselineskip=20pt
\pagestyle{myheadings}
\noindent
\centerline{{\large\bf Four-photon interference: a realizable experiment}}\\*[0.2cm] 
\centerline{{\large\bf to demonstrate violation of EPR postulates}}\\*[0.2cm]
\centerline{{\large\bf  for perfect correlations}}\\*[1.5cm]
\bigskip
\centerline{P. Hariharan${}^\dagger$, J. Samuel${}^*$ and Supurna Sinha}
\centerline{ Raman Research Institute, Bangalore 560 080,
India}
%\vspace{3cm}
\bigskip
\centerline{~~~~~~~~}
\centerline{~~~~~~~~}
\centerline{ Short title: Violation of EPR for perfect correlations}
\bigskip
\begin{abstract}

Bell's theorem reveals contradictions between the predictions 
of quantum
mechanics and the EPR postulates for a pair of particles only in situations
involving imperfect statistical correlations.
However, with three or more particles,
contradictions emerge 
even for perfect correlations. We describe an experiment 
which can be realized in the laboratory, using 
four-photon entangled states generated by parametric down-conversion,
to demonstrate this contradiction at the level of perfect correlations.

\end{abstract}
\vspace*{5cm}
\bigskip
{$\dagger$ Permanent address: School of Physics,
University of Sydney, Australia 2006}
\vspace*{.3cm}
{${}^*$ email: sam@rri.ernet.in
\newpage
\section{Introduction}
Einstein, Podolsky and Rosen \cite{EPR} (EPR) presented their 
famous {\it gedankenexperiment} in 1935 with the aim of showing that quantum
mechanics (QM) was not a complete description of physical
reality. A complete description, in their view, would require
the introduction of additional variables, usually referred to
as hidden variables. They outlined a program to reproduce the
predictions of QM using local hidden variable (LHV) theories.

This program was challenged by Bell \cite{Bell} in 
1964 when he proved that any hidden variable theory that incorporated the
concepts of locality and reality would be inconsistent with
certain predictions of QM. In particular, he showed \cite{Bell2}
that it was
possible to derive from the postulates of EPR  an inequality
which was violated by statistical predictions of QM for
a pair of particles. This violation has been observed in a
number of experiments involving interference of pairs of photons
produced in an entangled state [4-7].

A shortcoming of all these experiments is that, with a pair of
particles, Bell's theorem reveals contradictions between the
predictions of QM and EPR's postulates only in situations
involving imperfect statistical correlations: no contradictions
appear with perfect correlations.
A way to overcome this limitation, which was proposed by
Greenberger, Horne and Zeilinger \cite{GHZ} (GHZ) is to use three or more
particles in an entangled state.  
Greenberger {\it et al.} \cite{GHSZ} (GHSZ) 
have shown that, in this case, contradictions emerge even
at the level of perfect correlations. They also
described 
a {\it gedankenexperiment} using three entangled 
photons to illustrate this point.

Recent work suggests that it is possible to generate
entangled four-photon states by parametric down-conversion
of two pump photons \cite{Tewari}. 
Based on this work, we describe a design for a four-photon
interference experiment which can be realized in the laboratory.
We show how, with just two measurements, one can demonstrate that
quantum mechanics contradicts LHV theories, even at the level of perfect 
correlations.

\section{Generation of four photon entangled states}
Parametric down conversion in a crystal exhibiting a
$\chi^{(2)}$ nonlinearity makes it possible to convert a pump photon
(frequency $\omega_p$) into a pair of highly correlated photons
with frequencies $\omega_{d}{}^{(1)}$ and $\omega_{d}{}^{(2)}$, where 
$\omega_{d}{}^{(1)}+\omega_{d}{}^{(2)}=\omega_p$
[11-13].
These photons are generated almost simultaneously (within the correlation
time $\tau_d$ of the down-converted photons). 
While the frequency of each down-converted photon may
vary over an appreciable range, the sum of their frequencies is
fixed to within the pump bandwidth. The down-converted photons
are therefore described by an energy entangled state.

Recent work has shown that it should be possible to extend this
process to
generate entangled four-photon states from {\it two}
pump photons by achieving the required phase-matching conditions
in a non-linear crystal with two non-collinear pump beams \cite{Tewari}.
While the susceptibility for this two-photon down-conversion
process is low, the gain depends on the second power of the pump
amplitude, so that it should be possible to obtain an
appreciable yield by using pulsed pump beams with high peak power.
Further improvements in yield may be possible by the 
use of a resonant cavity (See Appendix D for further details
of the experiment).

In such an arrangement,
\begin{equation}\omega_d{}^{(1)}+\omega_d{}^{(2)}+\omega_d{}^{(3)}+\omega_d{}^{(4)}=
2\omega_p,\label{freqsum}\end{equation}
where $\omega_d{}^{(1)}$,..., $\omega_d{}^{(4)}$ are the frequencies of
the down-converted photons and $\omega_p$ is the frequency of the
pump photons, and 
\begin{equation}{\bf k}_d{}^{(1)}+{\bf k}_d{}^{(2)}+{\bf k}_d{}^{(3)}+{\bf k}_d{}^{(4)}=
{\bf k}_{p}{}^{(1)}+{\bf k}_{p}{}^{(2)},\label{wavevec}\end{equation}
where ${\bf k}_d{}^{(1)}$,..., ${\bf k}_d{}^{(4)}$ are the 
wave vectors of the down-converted
photons, and ${\bf k}_p{}^{(1)}$ and ${\bf k}_p{}^{(2)}$ are 
the wave vectors of the pump photons.
\section{Four photon interferometer}
The four-photon interferometer shown in figure 1 is an extension
of a two-photon interferometer described by Franson \cite{Franson}
in which each of the four down-converted photons enters one of four
interferometers, each with a short path ($s_j$) of length $S_j$ and  a
long path ($l_j$) of length $L_j$. The optical path difference $\Delta
L_j=L_j-S_j$ in each of the interferometers, which can be varied
by translating the right-angle prisms, 
is greater than the coherence length $c\tau_d$ of the down-converted
photons, so that no second-order interference effects due to
single photons are observed in the individual interferometers.

If, in a pair of interferometers, we consider the four processes leading
to photon counts 
($s_i-s_j, s_i-l_j, l_i-s_j, l_i-l_j$), and the difference of the optical
path differences ($\Delta L_{ij}=\Delta L_i-\Delta L_j$) is less
than the coherence length of the pump beam, the $l_i-l_j$ and $s_i-s_j$
processes are indistinguishable from each other. However, the
other two processes ($s_i-l_j$ and $l_i-s_j$) can be distinguished from the
$l_i-l_j$ and $s_i-s_j$ processes by the relative time lag of the
photons \cite{Kwiat}. It is then possible, with fast coincidence counters,
to reject counts arising from the $s_i-l_j$ and $l_i-s_j$ processes, so that
we are only concerned with coincidences due to the $l_1 -l_2- l_3-
l_4$ 
and $s_1- s_2 -s_3- s_4$ 
processes in the four interferometers (See Appendix D for further
comments on coincidence counts).

\section{Four photon interference}
  A four-photon event is recorded when photons are detected in coincidence
(within the detector response time) in all the four
interferometers. 
 Since the four photons are generated (almost) 
simultaneously,
such a coincidence could either be due to four photons which
all took the short path ($|s>_1|s>_2|s>_3|s>_4$) or the long path
$(|l>_1|l>_2|l>_3|l>_4)$. Following Feynman \cite{Feynman}, 
we can compute the amplitude (see refs. \cite{Kwiat} and \cite{Sorkin})
for the arrival of four coincident photons by 
summing the amplitudes for these indistinguishable
alternatives. We have 
\begin{equation}
|\Psi>= |s>_1|s>_2|s>_3|s>_4+ \exp(i\Phi)|l>_1|l>_2|l>_3|l>_4, \label{sup}
\end{equation}
where the relative phase $\Phi$ of the interfering 
$s_1-s_2-s_3-s_4$ and $l_1-l_2-l_3-l_4$ processes is the sum of the relative phases
acquired by the individual photons in the  four interferometers,
so that
\begin{equation}\Phi=\phi_1+\phi_2+\phi_3+\phi_4,\label{phasesum}\end{equation} 
where $\phi_i=(\omega_p/2c)\Delta L_i, (i=1,...,4)$ 
is the
phase difference between the beams traversing the two arms of
the $i$th interferometer.
The predicted coincidence count is obtained by squaring the
amplitude and is therefore proportional to 
\begin{equation}
|(1/2)(1+\exp(i\Phi))|^2=(1/2)(1+\cos(\Phi)),\label{main}
\end{equation}
where the constant
of proportionality includes the intensity of the source,
the detector efficiency (See Appendix C) and the losses in the system. 
A formal field-theoretic analysis which leads to the same result
is presented in Appendix A. If the detectors are as nearly alike
as possible, we can assume fair sampling, so that the number of coincidences
actually measured in any situation is proportional to those
expected for a perfect system.

As can be seen, QM predicts that the coincidence rate $R_c$ 
depends only on $\Phi$, the sum of the phase delays $\phi_i$
in the
four interferometers. The coincidences will be perfectly
correlated ($R_c=1$) when
$\Phi=0$ and perfectly anticorrelated ($R_c=0$) when $\Phi=\pi$. When 
$\Phi=0$, detection of a photon in three interferometers 
would imply the coincident
detection of a photon in the fourth interferometer. When $\Phi=\pi$, detection
of a photon in three of the interferometers 
would preclude the coincident detection
of a photon in the fourth interferometer.  
Let us define a parameter (analogous to the visibility
for sinusoidal fringes) 
\begin{equation}
{\cal Q}=\frac{R_c{}^{(0)}-R_c{}^{(\pi)}}{R_c{}^{(0)}+R_c{}^{(\pi)}} 
\label{vis}\end{equation}
whose value quantum mechanics predicts to be unity.
As we will see in the next section,
LHV theories cannot explain this value of ${\cal Q}$.

\section{LHV predictions}
It is convenient in discussing the four-photon interferometer to use
the language of spins traditionally used in the EPR literature. 
Traversals of the long and short arms of an interferometer are thought
of as basis states $|s>$ and $|l>$ correponding to ``spin up'' and
``spin down'' along the $z$ axis. 
The superposition of these states with a phase difference  $\phi$
\begin{equation}|\psi>=1/(\sqrt{2})(|s>+\exp(i\phi)|l>),\label{psisum}\end{equation}
 in spin language, is a state on the equator of the
Poincar\'e sphere of states of a spin-half particle where $|l>$
and $|s>$ are the North and South poles. The choice of a phase
delay $\phi_i$
in the $i$th interferometer corresponds to the choice
of a direction in the $x-y$ plane along which one measures
spin in Bohm's version \cite{Bo} of the EPR {\it gedankenexperiment}.

An LHV description \cite{GHZ} of the four photon interferometer
requires the use of a space $\Lambda$, the space of complete
states whose elements are written $\lambda$,
with a probability measure $\rho$. The expectation value of coincidence
counts is then
\begin{equation}\frac{1+E^\Psi(\phi_1,\phi_2,\phi_3,\phi_4)}{2},\label{exp}\end{equation}
where 
\begin{eqnarray}
E^\Psi(\phi_1,\phi_2,\phi_3,\phi_4)&=&<A(\phi_1)
B(\phi_2)C(\phi_3)D(\phi_4)>\nonumber\\
\mbox{  }&=&\int_{\Lambda}A_\lambda(\phi_1)
B_\lambda(\phi_2)C_\lambda(\phi_3)D_\lambda(\phi_4) d\rho,\label{expabcd}
\end{eqnarray}
and  $A_\lambda(\phi_1),
B_\lambda(\phi_2),C_\lambda(\phi_3),D_\lambda(\phi_4)$ are
four functions of $\lambda$ which take
values $\pm1$. Locality is built into the 
theory by the fact that $A_{\lambda}(\phi_1)$ 
is independent
of $\phi_2,\phi_3,\phi_4$, $B_{\lambda}(\phi_2)$ is 
independent of $\phi_1,\phi_3,\phi_4$,
and so on. 

\subsection{Perfect correlations: ideal experiment} 
Following GHZ \cite{GHZ,GHSZ} we can show that an LHV theory cannot 
reproduce the predictions of QM even at the level of perfect
correlations.\\ 

{\it Proof:} 
Let us suppose functions  $A_{\lambda}(\phi_1),B_{\lambda}(\phi_2),
C_{\lambda}(\phi_3),D_{\lambda}(\phi_4)$  exist, satisfying 
the relations

\begin{equation} <A(\phi_1)B(\phi_2)
C(\phi_3)D(\phi_4)>=1, \,\,{\rm for}\,\, \Phi=0,
\label{expone}\end{equation}
and
\begin{equation} <A(\phi_1)B(\phi_2)
C(\phi_3)D(\phi_4)>=-1, \,\,{\rm for}\,\, \Phi=\pi.
\label{exptwo}\end{equation}

Since the quantity in brackets can only take values $\pm1$, it follows that
everywhere in 
$\Lambda$ (except possibly for a set of measure zero),
\begin{equation} A_{\lambda}(\phi_1)B_{\lambda}(\phi_2)
C_{\lambda}(\phi_3)D_{\lambda}(\phi_4)=1 ,\,\,{\rm for}\,\,
\Phi=0,\label{one}\end{equation}
and
\begin{equation} A_{\lambda}(\phi_1)B_{\lambda}(\phi_2)
C_{\lambda}(\phi_3)D_{\lambda}(\phi_4)=-1 ,\,\,{\rm for}\,\, \Phi=\pi.
\label{two}\end{equation}
It then follows that 
\begin{equation} A_{\lambda}(-\phi) C_{\lambda}(\phi)
D_{\lambda}(0)B_{\lambda}(0)=1, \label{e1}\end{equation}
and
\begin{equation}B_{\lambda}(0) A_{\lambda}(-\phi)D_{\lambda}(\phi)
C_{\lambda}(0)=1. \label{e2}\end{equation}
Multiplying equations (\ref{e1}) and (\ref{e2}), 
we get, since $(A_{\lambda}(-\phi))^2=
(B_{\lambda}(0))^2=1$,
\begin{equation} C_{\lambda}(0)D_{\lambda}(0)
C_{\lambda}(\phi)D_{\lambda}(\phi)=1. \label{e3}\end{equation}
But 
\begin{equation} A_{\lambda}(0)B_{\lambda}(0)
C_{\lambda}(0)D_{\lambda}(0)=1. \label{e4}\end{equation}
Therefore
\begin{equation} A_{\lambda}(0)B_{\lambda}(0)
C_{\lambda}(\phi)D_{\lambda}(\phi)=1. \label{e5}\end{equation}
However, if we set $\phi=\pi/2$ in equation (\ref{e5}), 
it contradicts equation (\ref{two}). 
It follows that functions $A_{\lambda}(\phi_1),B_{\lambda}(\phi_2),
C_{\lambda}(\phi_3),D_{\lambda}(\phi_4)$
satisfying equations (\ref{one}) and (\ref{two}) do not exist. 
Accordingly, LHV theories cannot reproduce the
predictions of QM even at the level of perfect correlations. 

\subsection{Perfect correlations: real experiment} 
From  section  4 we see that QM predicts a ${\cal Q}$ of unity
in an
ideal experiment. 
However, in any real experiment,
one would obtain a value for ${\cal Q}$ less than unity because of
imperfections in the system. However, as  noted by Ryff \cite{Ryff},
``if a theorem is valid whenever we have perfect correlations, it
cannot be totally wrong in the case of almost perfect correlations''. 
We show below that with four-photon interference, 
a value of ${\cal Q}$  greater than 0.5 is
enough to
rule out LHV theories. We do this by going beyond the original argument
of GHZ  \cite{GHZ,GHSZ} to allow for 
experimental imperfections(${\cal Q}<1$). Mermin
 \cite{Mermin} has given an elegant and general analysis of the 
contradiction between quantum mechanics and LHV theories
for $n$ spin-$1/2$
particles in an entangled state and our bound on ${\cal Q}$ agrees with 
the restriction of Mermin's analysis to the case of four particles.

Given two functions $f$ and $g$ on $\Lambda$,
let us define an inner product (or cross correlation)
\begin{equation}<f g>=\int_{\Lambda}f_{\lambda} 
g_{\lambda} d\rho.\label{in}\end{equation}
We will only need to deal with functions which satisfy the condition
\begin{equation}
<ff>=1.
\label{norm}
\end{equation}
We then have the following lemma.\\

Lemma: Let $f,g,h$ be three functions on $\Lambda$ with values $\pm1$.
Then, \begin{equation}<fh>\geq <fg>+<gh>-1.\label{ineq}\end{equation} 
We present a proof and a geometrical interpretation of this lemma
in Appendix B.

Let us then suppose that there exist functions  
$A_{\lambda}(\phi_1),B_{\lambda}(\phi_2),
C_{\lambda}(\phi_3),D_{\lambda}(\phi_4)$   satisfying 
the relations
\begin{equation} <A(\phi_1)B(\phi_2)
C(\phi_3)D(\phi_4)>={\cal Q}, 
\,\,{\rm for}\,\, \Phi=0,\label{expvisone}\end{equation}
and 
\begin{equation} <A(\phi_1)B(\phi_2)
C(\phi_3)D(\phi_4)>=-{\cal Q} ,\,\,{\rm for}\,\, \Phi=\pi,\label{expvistwo}\end{equation}

where $0\leq{\cal Q}\leq1$. 
(Note that in the
limit, when ${\cal Q}\rightarrow 1$, we recover equations (10) and
(11).) We can no longer argue, as we
did before, that the angular brackets in equations (\ref{expone})
and (\ref{exptwo}) 
can be removed.
However, the lemma can be used to determine the
maximum allowed value for ${\cal Q}$.

From equation (\ref{expvisone}),  it follows that 

\begin{equation} <(A(-\phi) C(\phi)
D(0))(B(0))>={\cal Q}, \end{equation}
and
\begin{equation} <(B(0))(A(-\phi)D(\phi)
C(0))>={\cal Q}. \end{equation}
If we use the lemma, with 
\begin{eqnarray}
f&=&A(-\phi) C(\phi)D(0),\\ 
g&=&B(0),\\ 
h&=&A(-\phi)
D(\phi)
C(0),
\end{eqnarray}
and remember that  $(A_{\lambda}(-\phi))^2=
(B_{\lambda}(0))^2=1$,
 we get
\begin{equation} <C(0)D(0)
C(\phi)D(\phi)>\geq 2{\cal Q}-1. \end{equation}
However,
\begin{equation} <A(0)B(0)
C(0)D(0)>={\cal Q}, \end{equation}
so that, if we apply the lemma to these two relations, we find that
\begin{equation} <A(0)B(0)
C(\phi)D(\phi)>\geq 3{\cal Q}-2. \end{equation}
If then, we set $\phi=\pi/2$ and use equation (\ref{expvistwo}), we find that 
\begin{equation}
(-{\cal Q})\geq3{\cal Q}-2,\end{equation}
from which it follows that
\begin{equation}{\cal Q}\leq 1/2.\end{equation}
This result proves that LHV theories cannot yield a value of ${\cal Q}$
greater than 0.5.

Note that in our adaptation of the original GHZ argument \cite{GHZ,GHSZ},
it is not necessary to set the individual phase differences
$\phi_1,...,\phi_4$ to $0$ (or, more correctly, $2 m \pi$): it
is only necessary to set 
$\Phi$, the sum of these phase differences, 
to $0$ (or $2 m \pi$). In practice, it is difficult
(nearly impossible) to set the individual phase differences to
any preassigned value, since the optical path differences
in the individual interferometers are greater than the coherence
lengths of the down-converted photons; our adaptation eliminates
this problem and makes the experiment feasible.

In the actual experiment, one of the four interferometers
is adjusted initially so that the coincidence rate is a maximum.
The first measurement therefore corresponds to the condition
$\Phi=2m\pi$. 
A phase shift
of $\pi$ is then introduced in any one of the interferometers and the
event rate is measured at the resulting minimum. (Note that the introduction
of a further phase shift of $\pi$ in any 
of the interferometers would bring the 
event rate back to a maximum). The results of these two measurements
are inserted in equation (6) to obtain the value of the quantity 
${\cal Q}$. Any value greater than $0.5$ represents a breakdown
of LHV theories under perfect correlations.

The only data used correspond effectively to
values of $\Phi$ of $0$ (eq. 22) and $\pi$ (eq. 23). This is very much in
the spirit of the original GHZ argument \cite{GHZ,GHSZ}, 
which relies only on perfect 
correlations.
\subsection{Statistical correlations}
This experiment also makes it possible to demonstrate
violations of the original
Bell inequality (which uses statistical correlations
for two particles) in systems of four particles.
In this case, it is not necessary to adjust the value of $\Phi$
for the first measurement so that the coincidence rate
is a maximum; $\Phi$ can have any arbitrary value (say) $\Phi_0$.
Three more measurements
of the event rate are then made after introducing phase shifts of $\pi/2$,
successively, in three of the interferometers. We then have four values of
the event rate corresponding to values of $\Phi$ of $\Phi_0,\Phi_0+\pi/2,
\Phi_0+\pi$ and $\Phi_0+3 \pi/2$.

If one assumes that the fringe profile is sinusoidal, one can easily
determine the fringe visibility from these four measurements. Whereas
quantum mechanics predicts a visibility of unity, it has been shown 
\cite{Roy,Arde,Zu} that LHV theories cannot explain a fringe visibility
greater than $1/(2\sqrt{2})$. In this respect, a four-photon 
experiment offers a more probing test 
than three-photon experiments
for which the critical visibility is $1/2$. However, as explained
in \cite{Arde} this lower value for the critical visibility 
relies on the use of
{\it statistical correlations} rather than perfect correlations.

\section{Conclusion}
While most theoretical studies related to EPR 
have involved spin-($1/2)$ particles, 
actual experiments have used
optical analogs of such systems. In particular, 
all interferometric tests of Bell's 
inequality carried out so far have used entangled 
two-photon states [4-7]. In this case, LHV theories do not 
contradict QM at the level of perfect correlations. Therefore, tests of
LHV theories with two-photon states 
require measurements of statistical correlations.
EPR experiments involving more than
two particles (as in section 5.3) utilize extensions
of Bell's inequality and, therefore, also involve
statistical correlations. On the other hand, tests such as those
described in section 5.2 are based on the GHZ analysis \cite{GHZ,GHSZ}
and, therefore,
only involve perfect correlations.
Since perfect correlations formed
the basis of the original EPR criterion for ``elements 
of reality'', the contradiction emerging from the GHZ analysis \cite{GHZ,GHSZ}
strikes at the heart of the EPR program. 
We have described a realizable experiment involving four-photon interference
which demonstrates the
conflict between EPR and QM even at the level of
perfect correlations.

\section*{Acknowledgements} One of the authors (P.H) thanks the
International Centre for Theoretical Physics for support
under their Visiting Scholar Program.
\newpage
\section*{Appendix A}

The field theoretic analysis presented by Franson \cite{Franson} 
can be easily extended to the case of four-photon interference.
A field theoretic description has the advantage that it
is manifestly local. We sketch the
main ideas below using his notation.

We need only deal with scalar fields since the polarization is fixed 
throughout.
The scalar field operator $\psi({\vec r},t)$ is expanded in free space modes
as
\begin{equation}\psi({\vec r},t)=\sum_{\vec k} \frac{a_{\vec k}}{\sqrt{V}}\exp(i({\vec k}.{\vec r}-\omega t)).
\label{a1}\end{equation}
The time evolution of this operator is governed by the free
Hamiltonian of the electromagnetic field
and since \begin{equation}\psi(x+c\Delta t,t)=\psi(x,t-\Delta t),\label{a2}\end{equation}
the particle it describes moves at the speed of light. 

The field operator at the detector of the 
$i$th interferometer with the beam splitter
removed is given by $\psi_{0}({\vec r}_i,t)$. For 
each pair $(i,j)$ of the interferometers,
these operators 
satisfy the condition 
\begin{equation}\psi_{0}({\vec r}_i,t)\psi_{0}({\vec r}_j,t\pm\Delta t)|0>=0,
\label{a3}\end{equation}
which is analogous to
Franson's equation (5). With the beam splitter inserted in the interferometer,
the field operator at the $i$th detector becomes
\begin{equation}\psi({\vec r}_i,t)=(1/2)(\psi_0({\vec r}_i,t)+\exp(i\phi_i)\psi_0({\vec r}_i,t-\Delta t)),\label{a4}\end{equation}
where $i=1,...,4$.
The coincidence rate $R_c$ for the four detectors $D_1,D_2,D_3,D_4$,
with the beam splitters inserted, is then
\newpage
\begin{eqnarray}
R_c&=&\eta_1\eta_2\eta_3\eta_4 \times\nonumber \\
& &<0|\psi^{\dagger}(r_1,t)
\psi^{\dagger}({\vec r}_2,t)\psi^{\dagger}({\vec r}_3,t)
\psi^{\dagger}({\vec r}_4,t)
\psi({\vec r}_1,t)\psi({\vec r}_2,t)\psi({\vec r}_3,t)\psi({\vec
r}_4,t)|0>,\nonumber \\
\label{a5}\end{eqnarray}
where $\eta_i$ is the efficiency of the $i$th detector. Substituting
(\ref{a4}) in (\ref{a5}) and using (\ref{a3}), we obtain the result
\begin{equation}R_c=
\left(\frac{R_{c0}}{2^6}\right) \left(\frac{1+\cos(\Phi)}{2}\right),
\label{a6}\end{equation}
where 
\begin{eqnarray}R_{c0}&=&\eta_1\eta_2\eta_3\eta_4\times\nonumber \\
& &<0|\psi_0{}^{\dagger}({\vec r}_1,t)
\psi_0{}^{\dagger}({\vec r}_2,t)\psi_0{}^{\dagger}({\vec r}_3,t)
\psi_0{}^{\dagger}({\vec r}_4,t)
\psi_0({\vec r}_1,t)\psi_0({\vec r}_2,t)
\psi_0({\vec r}_3,t)\psi_0({\vec r}_4,t)|0>\nonumber \\
\label{a7}
\end{eqnarray}
and $\Phi=\phi_1+\phi_2+\phi_3+\phi_4$.
\section*{Appendix B}
Let us consider three real-valued 
functions $f,g,h$ on $\Lambda$. One can think of these
functions as elements of a vector space. 
We only need to work in a three dimensional
subspace containing $f,g,h$. 
If $f,g,h$ are unit vectors, $<f g>=\cos(\theta_{fg})
$, $<g h>=\cos(\theta_{gh})$ and $<f h>=\cos(\theta_{fh}),$ 
where the angles $\theta_{fg},\theta_{gh},\theta_{fh}$
are defined to be less than $\pi$ and represent the angles between the unit
vectors $f,g,h$. These angles can also be interpreted as the lengths
of the shortest geodesics between the tips of the vectors $f,g,h$ 
on the unit sphere. From the triangle inequality, it then follows that
\begin{equation}\theta_{fh}\leq\theta_{fg}+\theta_{gh}.\label{b1}\end{equation}
Since $\cos(\theta)$ is a decreasing function of its argument for $0\leq
\theta \leq \pi$, we arrive at the result
\begin{equation}\arccos(<f h>)\geq\arccos(<f g>)+\arccos(<g h>).\label{b2}\end{equation}
This inequality has a clear interpretation in terms of the triangle
inequality on the unit sphere. 

The inequality we use in the text is similar in spirit. We 
only need to deal with 
functions $f,g,h$ which take values 
$\pm1$ and, in this case, can make
a stronger statement.
(Such functions and their correlations are of interest in
digital signal processing, in communication theory \cite{Vleck}
and radio astronomy \cite{Weinreb}). We then have
\begin{equation}<f g>=2 \Omega(fg)-1,\label{b3}\end{equation}
where $\Omega(fg)$ is the volume of the domain ${\cal D}(fg)$ 
in $\Lambda$ where $f$ and $g$ agree.
Similarly, $<g h>=2 \Omega(gh)-1$
and $<f h>=2 \Omega(f h)-1$. Since the domain of agreement
${\cal D}(fh)$ between
$f$ and $h$ includes at least ${\cal D}(fg)\cap{\cal D}(gh)$,
the intersection of 
the domains of agreement between $f$ and $g$ (${\cal D}(fg)$) and
between $g$ and $h$ (${\cal D}(gh))$,
we conclude that
\begin{equation}\Omega(fh)\geq\Omega(fg)+\Omega(gh)-1.\label{b4}\end{equation}
It follows immediately that 
\begin{equation}<fh>\geq <fg>+<gh>-1\label{b5}\end{equation}. 

It is worth noting that if we write $F(x)=(1-x)/2$, inequality
(45) reads 
\begin{equation}F(<fh>)\leq F(<fg>)+F(<gh>),\label{b6}\end{equation}
which is similar to inequality (42), with $F(x)$ replacing the function
$\arccos(x)$. Both these inequalities express the idea that if 
$f$ and $g$ are highly correlated and $g$ and $h$ are highly 
correlated, then $f$ and $h$ must be correlated to some extent.

\section*{Appendix C:  Detector efficiency}

In avalanche photodiodes, the only significant loss mechanism is reflection of
the incident photons.  The detection efficiency is therefore given by the
relation
\begin{equation}
\eta = 1 - R,
\end{equation}
where $R$ is the fraction of photons reflected at the surface of the
photocathode.  It is, therefore,  possible to reduce the loss due to this
cause to negligible levels by using a number of photodiodes in a light-
trapping arrangement \cite{zal}.  

A  simple trap-detector which can be used for photon counting uses only two
photodiodes \cite{har}.  In this arrangement, as shown in Fig.$ 2$, 
the incident beam
undergoes three reflections at the photodiodes before exiting.  The fraction
of the photons lost by reflection is then
\begin{equation}
R_{2} =  R^{3},
\end{equation}
and summation of the outputs of the two photodiodes should yield a detection
efficiency
\begin{equation}
\eta_{2} = 1 - R^{3}.
\end{equation}
With commercial avalanche photodiodes,  for which $R$ is typically around
$0.3$,   it should be possible to obtain an increase in detection efficiency
from $70\%$ to $97\%$. 
\section*{Appendix D:  Generation of four-photon states}

In the usual parametric process, yielding two down-converted photons,  a 25 mm
long ADP crystal pumped by a 9 mW He-Cd laser ($\lambda = 325 nm$), yields, at
a 2 mm aperture placed at a distance of 1 m from the crystal, a down-converted
flux of  $4 \times 10^{5}$ photons / second, for each beam \cite{burn}, corresponding
to a down-conversion efficiency of $3  \times 10^{-11}$.  However, crystals
such as beta-barium borate (BBO) are now available with a nonlinear
coefficient 5 times higher than ADP.  In addition, it should be possible to
obtain an increase in down-conversion efficiency by placing the crystal in a
short resonant cavity \cite{tew}.  If we use a 1.5 cm long BBO crystal, placed in a
short cavity with mirrors whose reflectivity is chosen so that the effective
length of the crystal is increased to around 7.5 cm,  it should be possible to
obtain a down -conversion efficiency of $4.5 \times 10^{-10}$.

The nonlinear susceptibility involved in the production of the four-photon
field is, to a first approximation, the square of the nonlinear susceptibility
involved in the production of two down-converted photons \cite{tew}, so that the
down-conversion efficiency, in this case, would work out to $2 \times
10^{-19}$.  However, with two pump beams, the gain depends on the second power
of the pump amplitude \cite{tew}.  As a result, the output with pulsed pump beams
with high peak power can be several orders of magnitude greater than that
obtained with continuous-wave excitation at the same average power.  
   
With a laser generating pulses with a duration of 1 $\mu$s, at a repetition
rate of 10 pulses/second, it should be possible to obtain a peak power that is
$10^5$ times greater than the average power, and an improvement in down-
conversion efficiency by a factor of this order.  Accordingly, with an average
power of 100 mW (corresponding to a peak power of 10 kW), it should be
possible to obtain a total down-converted flux in the four output beams of $4
\times 10^3$ photons/second, or $10^3$ photons/second in each beam.  After
allowing for losses, it should be possible to obtain a flux of 100
photons/second at the output from each of the four interferometers, which
should permit useful measurements.    

The use of a pulsed pump beam might be expected, at first sight, to create
problems connected with the spectral coherence of the pump beams and the time
resolution of the detectors.  However, with a pulse duration of 1 $\mu$s, the
coherence length of the pump beams, with a properly designed laser cavity,
would be greater than 100 m.  On the other hand, the coherence length of the
down-converted beams, which would be determined by the decay time of the
cavity modes (in this case, about 0.5 ns), would be less than  0.15 m.
Accordingly, it would be possible to avoid second-order interference fringes
by working with an optical path difference greater than this value, without a
significant loss in the visibility of fourth-order interference effects. 

As mentioned earlier, with a crystal placed in a resonant cavity, the light
beams have an intrinsic bandwidth determined by the bandwidth of the cavity.
This is consistent with the picture that the four down-converted photons are
produced simultaneously, but then escape independently within a time interval
equal to the decay time of the cavity \cite{coll,sand}, 
which, as mentioned earlier, is
around 0.5 ns.  Since this time interval is much less than the time resolution
of a fast photodetector (say, 1.5 ns), the effects of such a deviation from
simultaneity would not be noticeable.  

Finally, we need to consider the probability of accidental coincidences.
Since the output from each interferometer consists of a series of pulses with
a duration of 1 $\mu$s, each containing about 10 photons, the probability of
detecting a single photon in a time window of 1.5 ns would be 0.015.  The
ratio of the probability of accidental coincidences at the four outputs, due
to uncorrelated photons, to that for actual coincidences would be only
marginally higher, at around $0.02$.  This proportion of accidental coincidences
should not have a significant effect on the visibility of fourth-order
interference effects produced by the four down-converted beams.

\newpage
\section*{Figure Captions}
Fig. 1. Schematic of the four-photon interferometer.
\vskip .5cm 
\noindent Fig. 2. Optical configuration for a single-photon trap detector using
two avalanche photodiodes.
\newpage
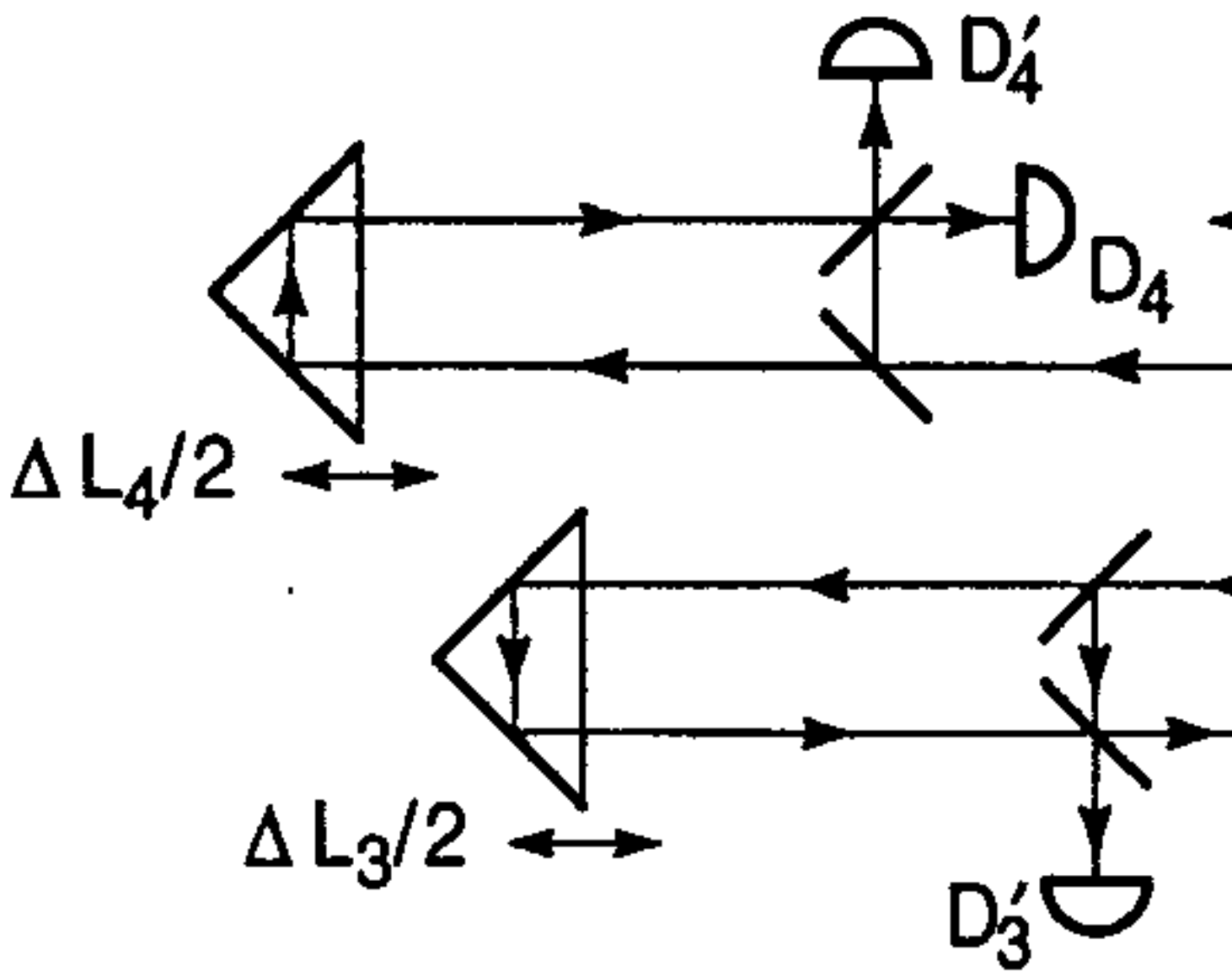
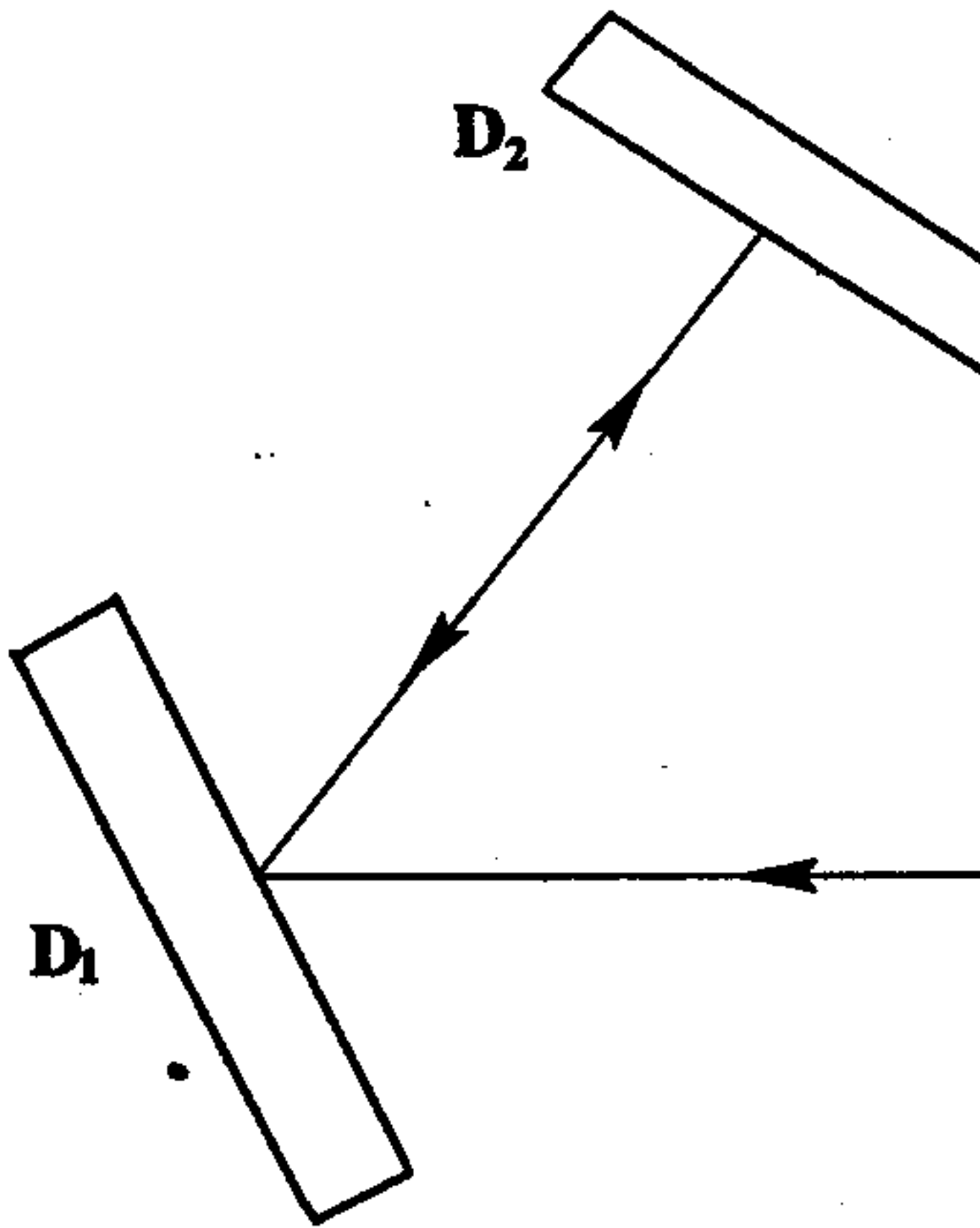
\end{document}